\documentstyle[12pt,epsf]{article}
\epsfverbosetrue
\def\figlabel#1{\xdef#1{\thefigure}}
\def\fig#1{fig.~#1}

\def\figalign#1#2#3#4#5#6{
\begin{figure}
\centerline{
\hbox to 2.5truein{\vtop{\hsize=2.5truein\epsfxsize=6cm
\centerline{\epsfbox{#1} }
\caption[]{#3}
\figlabel{#2}
}}
\qquad\hbox to 2.5truein{\vtop{\hsize=2.5truein\epsfxsize=6cm
\centerline{\epsfbox{#4} }
\caption[]{#6}
\figlabel{#5}
}}
}
\end{figure}
}

\def\be{\begin{equation}}
\def\ee{\end{equation}}
\def\bea{\begin{eqnarray}}
\def\eea{\end{eqnarray}}

\begin{document}
\begin{titlepage}
\begin{flushright}
{ ~}\vskip -1in
CERN-TH/96-166\\
US-FT-34/96\\
\date\\
hep-th/9606191
\end{flushright}
\bigskip
\begin{center}
{\LARGE MORE ON SOFTLY BROKEN \\
   \bigskip
$N=2$ QCD}
\vskip 0.9truecm

{Luis \'Alvarez-Gaum\'e$^{a}$
\\and\\ 
Marcos Mari\~no$^{a,b}$}

\vspace{1pc}

{\em $^a$ Theory Division, CERN,\\
 1211 Geneva 23, Switzerland.\\
 \bigskip
  $^b$ Departamento de F\'\i sica de
Part\'\i culas,\\ Universidade de Santiago
de Compostela,\\ E-15706 Santiago de Compostela, Spain.}\\

\vspace{5pc}

{\large \bf Abstract}
\end{center}

We extend previous work on the soft breaking of $N=2$ supersymmetric 
QCD. We present the formalism for the breaking due to a dilaton 
spurion for a general gauge group and obtain the exact effective potential. 
We obtain some general features of the vacuum structure in the pure 
$SU(N)$ Yang-Mills theory and we also
derive a  general mass formula for this 
class of theories, in particular we present explicit results for the 
mass spectrum in the $SU(2)$ case. Finally we analyze the vacuum 
structure of the $SU(2)$ theory with one massless hypermultiplet. This theory 
presents dyon condensation and a first order phase transition in the 
supersymmetry breaking parameter driven by non-mutually local BPS 
states. This could be a hint of Argyres-Douglas-like phases in 
non-supersymmetric gauge theories.

\vfill
\vspace{5pc}
\begin{flushleft}
{ ~}\vskip -1in
CERN-TH/96-166\\
June 1996
\end{flushleft}

\end{titlepage}

\def\theequation{\thesection.\arabic{equation}}

\section{Introduction and conclusions}
\setcounter{equation}{0}
Recently there has been great progress in understanding 
the dynamics of supersymmetric gauge theories in four dimensions. 
For $N=1$ theories exact results have been obtained
\cite{none} using the holomorphy
properties
of the superpotential and the gauge kinetic function,
culminating in Seiberg's non-abelian
duality conjecture \cite{nadual}.
In two remarkable papers \cite{swone,swtwo},
 Seiberg and Witten obtained
 exact information on $N=2$ Yang-Mills theories with gauge group
$SU(2)$ and $N_{f} \le 4$
 flavour multiplets. Their work was extended to
other groups in \cite{kl, klt, ds, groups}.
 One of the crucial advantages of using
$N=2$ supersymmetry is that the low-energy
 effective action in the Coulomb phase up to two derivatives
 ({\it i.e.} the K\"ahler potential,
the superpotential and the gauge kinetic function in
$N=1$ superspace language)
 are determined in terms of a single holomorphic function called the
 prepotential \cite{four}, which was exactly determined 
in \cite{swone,swtwo} using
some plausible assumptions and many consistency conditions.
For $SU(2)$ the solution is neatly presented by associating
to each case an elliptic curve together with
a meromorphic differential
of the second kind whose periods completely
determine the prepotential.
For other gauge groups \cite{kl, groups} the solution
is again presented in terms
of the period integrals of a meromorphic differential
on a Riemann surface
whose genus is the rank of the group considered.
It was also shown in \cite{swone,swtwo} that by soft breaking $N=2$
down to $N=1$ (by adding a mass term for the adjoint $N=1$ chiral
multiplet in the $N=2$ vector multiplet) confinement follows
due to monopole condensation \cite{five}.

With all this new information it is also tempting to analyze the 
dynamics of non-supersymmetric gauge theories in order 
to determine to what extent these results depend on the 
supersymmetry structure, and perhaps to obtain 
exact information about ordinary QCD.
 A useful avenue to explore is soft supersymmetry
breaking. The structure of soft supersymmetry breaking in $N=1$ theories has
been known for some time \cite{girar}. In \cite{softone,softwo}
 soft breaking terms are used to explore $N=1$ supersymmetric QCD
(SQCD)
 with gauge group $SU(N_c)$ and $N_f$ flavours of quarks,
 and to extrapolate the exact results in \cite{none}
concerning the superpotential and the phase structure
 of these theories in the absence of supersymmetry.
 This leads to expected and unexpected predictions for
 non-supersymmetric theories which may eventually be accessible to
lattice computations. Since the methods of \cite{swone,swtwo} 
provide us with
 the effective action up to two derivatives,
 the possibility was explored in \cite{soft}
 of breaking $N=2$ SQCD directly to $N=0$ preserving 
the analyticity properties of the Seiberg-Witten solution.
It has been shown in \cite{soft} that a natural way 
to accomplish this task is, essentially, to make the dynamical scale
 $\Lambda$ of the $N=2$ theory a function of an
$N=2$ vector multiplet. This multiplet is then frozen
to become a spurion whose $F$ and $D$-components break softly
$N=2$ down to $N=0$, in a way compatible with the
Seiberg-Witten monodromies. The spurion can be interpreted in terms 
of the string derivation of the
Seiberg-Witten solution in \cite{XI,XII}, based on
type II-heterotic duality. In this spirit, the soft breaking terms 
were rederived in \cite{soft} starting from
 the theory coupled to $N=2$ supergravity 
with a simple superpotential which breaks spontaneously
supersymmetry. The resulting theory at $N=0$ has a more 
restricted structure than those used in \cite{softone,softwo}, 
and it is possible to compute the exact effective 
potential. The $SU(2)$ case with $N_f=0$ and $N_f=2$ massless 
quark hypermultiplets were analyzed in detail, and it turned out that 
these softly broken theories have 
a unique ground state
(near the massless monopole points of
\cite{swone,swtwo}) 
and confinement occurs through monopole condensation. 
 
In this paper we extend the results of \cite{soft} in three 
different directions. First of all, we present the
 general formalism
for the breaking of supersymmetry due to a dilaton
spurion for a general gauge group with massless 
hypermultiplets, and we study  
the symplectic transformations of the various
quantities involved. The results agree with the general 
structure derived in \cite{XIII}
concerning the modification of the symplectic
transformations of special geometry in the
presence of background $N=2$ vector superfields. 
We also obtain the exact effective potential for the Coulomb 
phase in this 
general setting, and we derive some general features of 
the vacuum structure in the $SU(N)$ case studying the
 theory around the $N=1$ points described in \cite{ds}. 
Another important issue addressed in this paper is the mass spectrum 
of the softly broken theories. We prove a general mass formula 
for this class of theories, stating that the graded trace of the 
squared mass matrix is zero, as it happens in supersymmetric theories 
and also in some restricted models of $N=1$ soft supersymmetry breaking
 \cite{mass}. We also obtain explicit results for the 
mass spectrum in the case of $SU(2)$ theories. 
Finally, we consider the softly broken $SU(2)$ theory with one 
massless hypermultiplet. The vacuum structure of this model is 
very interesting, as it presents two degenerate ground states
with dyon condensation. The corresponding BPS states are not 
mutually local 
and as the supersymmetry breaking parameter is turned on the 
regions where these states get a VEV begin to overlap. The 
possibility of this kind of behaviour was pointed out in \cite{soft}, 
and here we find an explicit realization. We argue that the 
effective potential description of this situation is still reliable 
as long as the condensates do not attain the singularities associated 
to the other states, and this allows us to obtain some hints of the 
dynamics involving non-local objects. Within this range of validity 
we find a first order phase transition to a new ground state 
located in the region where the two condensates overlap. 
Although the interpretation of this new vacuum is not clear to us, 
it may correspond to a new kind of phase similar to the Argyres-Douglas phase
\cite{ad}. The dynamics of this phase transition looks very much like the 
phase transitions in the theta angle leading to an oblique confinemente phase. 
Other possible interpretation would be the formation of a bound state 
condensing in the new vacuum. 

We see that the study of $N=2$ soft supersymmetry breaking 
gives a rich variety of dynamical structures. We believe that 
the introduction of bare masses for the quark hypermultiplets will also 
lead to interesting field theoretic phenomena that may shed some light 
on the exact dependence on the quark masses of the effective Goldstone boson 
lagrangian describing chiral symmetry breaking. 
This situation is presently under study.   

The organization of this paper is as follows:
In section two we extend the formalism of \cite{soft} 
to arbitrary gauge groups with massless hypermultiplets.           
In section three we derive the exact effective potential 
in the Coulomb phase and the basic tools to 
analyze the vacuum structure. In section four 
we analyze in some detail the case of the $SU(N)$ Yang-Mills theory 
without hypermultiplets. In section five
we present a general mass sum rule, and also obtain 
explicit results for the masses
in the $SU(2)$ case. In section six, we consider the $SU(2)$ 
theory with one massless hypermultiplet.

\section{Breaking $N=2$ with a dilaton spurion: general gauge group}
\setcounter{equation}{0}
In this section we present the generalization of 
the procedure introduced 
in \cite{soft} to $N=2$ Yang-Mills theories 
with a general gauge group $G$ of rank 
$r$ and massless matter hypermultiplets. 

The low energy theory description of the 
Coulomb phase \cite{swone} involves $r$ abelian 
$N=2$ vector superfields $A^i$, $i=1, \cdots, r$ 
corresponding to the unbroken 
gauge group ${U(1)}^r$. The holomorphic prepotential 
${\cal{F}} (A^i,\Lambda)$
 depends on 
the $r$ superfields $A^i$ and the dynamically 
generated scale of the theory, 
$\Lambda$. The low energy effective lagrangian 
takes the form (in $N=1$ notation)
\cite{swone}:
\be
{\cal L}={1 \over 4 \pi}{\rm Im}
\Bigl[ \int d^4 \theta {\partial {\cal F}\over
\partial A^i} {\overline A}^i + {1\over 2}
\int d^2 \theta {\partial ^2 {\cal F} \over
\partial A^i \partial A^j} W^i_{\alpha}W^{\alpha  j} \Bigr],
\label{lagr}
\ee
We define the dual variables, as in the $SU(2)$ case, by
\be
a_{D,i} \equiv {\partial {\cal F} \over \partial a^i}.
\label{two}
\ee
The K\"ahler potential and effective couplings 
associated to (\ref{lagr}) are:    
$$
K(a, {\bar a})={1 \over 4 \pi}{\rm Im} a_{D,i}{\bar a}^i,
$$
\be
\tau_{ij}= {\partial ^2 {\cal F} \over \partial a^i
\partial a^j}, 
\label{kahler}
\ee
and the metric of the moduli space is given accordingly by:
\be
(ds)^2={\rm Im}{\partial ^2 {\cal F} \over \partial a^i
\partial a^j}da^i d{\overline a}^j.
\label{metric}
\ee 
We introduce now a complex space ${\bf C}^{2r}$ with elements of the form
\be
v=\left(\begin{array}{c}a_{D,i} \\
                    a^i \end{array}\right).
\label{vector}
\ee
The metric (\ref{metric}) can then be written as
\bea
(ds)^2 &=& -{i \over 2}\sum_{i} (da_{D,i}d{\overline a}^i
-d {\overline a}_{D,i}da^i)
\nonumber\\
& \nonumber \\
&=& -{i \over 2} \left(\matrix{da_{D,i}& da^i}\right)
\left(\matrix{0& {\bf 1} \cr
               -{\bf 1}&0\cr}\right) 
\left(\begin{array}{c}d{\overline a}_{D,i} \\
                    d{\overline a}^i \end{array}\right),
\label{metmat}
\eea
which shows that the transformations of $v$ 
preserving the form of the metric 
are matrices $\Gamma \in Sp(2r, {\bf Z})$. 
They verify $\Gamma^{\rm T} \Omega 
\Gamma=\Omega$, 
where $\Omega$ is the $2r \times 2r$ matrix 
appearing in (\ref{metmat}), and
can be written as:
\be
\left(\matrix{A& B \cr
               C&D\cr}\right) 
\label{simp}
\ee
where the $r \times r$ matrices $A$, $B$, $C$, $D$ satisfy:
\be
A^{\rm T}D-C^{\rm T}B= {\bf 1}_r, \,\,\,\ 
A^{\rm T}C=C^{\rm T}A, \,\,\,\ 
B^{\rm T}D=D^{\rm T}B.
\label{simple}
\ee
The vector $v$ transforms then as:
\be
\left(\begin{array}{c}a_{D} \\
                    a \end{array}\right) \rightarrow \Gamma
\left(\begin{array}{c}a_D\\
               a \end{array}\right) =\left(\begin{array}{c}Aa_D+Ba\\
               Ca_D+Da \end{array}\right).
\label{trans}
\ee
From this we can obtain the modular transformation 
properties of the prepotential
 ${\cal F}(a^i)$ (see \cite{sonn}). Since
\bea
{\partial {\cal F}_{\Gamma} \over \partial a^k} &=&
{\partial a^i_{\Gamma} \over \partial a^k}
{\partial {\cal F}_{\Gamma}\over \partial a^i_{\Gamma}}=
\Bigl( C^{ip} \tau_{pk}+D^i_k \Bigr)
\Bigl(A_i^j a_{D,j} +B_{ij}a^j \Bigr)\nonumber \\
& \nonumber \\
&=& (D^{\rm T} B)_{kj}a^j+(D^{\rm T}A)_k^j{ \partial {\cal F} \over \partial 
a^j} + 
(C^{\rm T} B)^p_j {\partial a_{D,p} \over \partial a^k} a^j\nonumber\\
& \nonumber\\
&+& (C^{\rm T}A)^{pj} {\partial a_{D,p} \over \partial a^k}a_{D,j},
 \label{xxiii}
\eea
using the properties (\ref{simple}) of the 
symplectic matrices we can integrate 
(\ref{xxiii}) to obtain:
\bea
{\cal F}_{\Gamma}&=& {\cal F} +
{1 \over 2} a^k(D^{\rm T} B)_{kj}a^j+ {1 \over 2} a_{D,k} 
(C^{\rm T}A)^{pj}a_{D,j}\nonumber\\
& \nonumber\\
&+& a^k (B^{\rm T}C)_k^j a_{D,j}.
\label{premono}
\eea
Starting with (\ref{premono}) we can prove that the quantity 
${\cal F}-{1/2}\sum_{i}a^ia_{D,i}$ 
is a monodromy invariant, 
and evaluating it asymptotically, one obtains the relation 
\cite{matone,sonn,ey}:
\be
{\cal F}-{1 \over 2}\sum_{i}a^i a_{Di}=-4 \pi i b_1 u,
\label{mato}
\ee
where $b_1$ is the coefficient of the 
one-loop $\beta$-function (for $SU(N_c)$ with $N_f$ 
hypermultiplets in the fundamental representation, 
$b_1=(2N_c-N_f)/16 {\pi}^2$) and 
$u=\langle {\rm Tr} \phi^2 \rangle$. With the normalization 
for the electric charge used in \cite{swtwo} and 
\cite{ds}, the r.h.s. of 
(\ref{mato}) is $-2 \pi i b_1 u$. 

As in the $SU(2)$ case, presented in \cite{soft}, 
we break $N=2$ supersymmetry down to 
$N=0$ by making the dynamical scale $\Lambda$ a 
function of a background vector 
superfield $S$, $\Lambda = {\rm e}^{iS}$. 
This must be done in such a way 
that $s$, $s_D= \partial {\cal F} /\partial s$ 
be monodromy invariant. To see this, 
we will derive a series of relations analogous 
to the ones in the $SU(2)$ case, 
starting with the following expression for the 
prepotential in terms of local 
coordinates:
\be
{\cal F}=\sum_{ij}a^i a^j f_{ij}(a^l/\Lambda),
\label{expansion}
\ee
where we take $f_{ij}=f_{ji}$. We define 
now a $(r+1)\times (r+1)$ matrix of couplings 
including the dilaton spurion $a^0=s$:
\be
\tau_{\alpha \beta}={{\partial}^2 
{\cal F}\over \partial a^{\alpha} a^{\beta}}.
\label{dilcouplings}
\ee
Greek indices $\alpha$, $\beta$ go from $0$ to $r$, 
and latin indices $i$, $j$ from
 $1$ to $r$. We obtain:
$$
a_{D,k}=2\sum_i a^if_{ik} + {1 \over \Lambda} \sum_{ij}a^i a^j f_{ij,k},
\nonumber
$$
$$
\tau_{ij}=2f_{ij}+{2 \over \Lambda}\sum_ka^k(f_{ik,j}+f_{jk,i})+
{1 \over \Lambda^2} a^ka^lf_{kl,ij},
\nonumber
$$
$$
\tau_{0i}=-{i\over \Lambda}\sum_{jk}a^ja^k(2f_{ij,k}+
f_{jk,i})-{i \over {\Lambda}^2}
\sum_{jkl}a^ja^ka^lf_{jk,li},
\nonumber
$$
\be
\tau_{00}=-{1 \over \Lambda}\sum_{ijk}a^ia^ja^kf_{ij,k}-
{1\over \Lambda^2}\sum_{ijkl}a^ia^ja^ka^lf_{ij,kl},
\label{sumas}
\ee
and the dual spurion field is given by:
\be
s_{D}={\partial {\cal F} \over \partial s}=
-{i \over \Lambda} \sum_{ijk}a^ia^jf_{ij,k}
\label{esdual}
\ee
The equations (\ref{sumas}) and (\ref{esdual}) give the useful relations:
$$
\tau_{0i}=i(a_{D,i}-\sum_j a^j\tau_{ji}), \,\,\,\,\,\ 
{\partial \tau_{0i} \over \partial a^k}=
-i\sum_j a^j {\partial \tau_{ij} \over 
\partial a^k},
$$
\be
{\partial \tau_{00} \over \partial a^k}=i\tau_{0k}-\sum_{ij}a^i a^j 
{\partial \tau_{ij} \over
\partial a^k}.
\label{rela}
\ee
Using now (\ref{mato}) one can prove that $s_D$ is a monodromy invariant,
\be
{\partial {\cal F} \over \partial s}=
i\Bigl(2 {\cal F} -\sum_i a^i a_{D,i}\Bigr)=
8 \pi b_1 u
\label{espurion}
\ee
and from (\ref{rela}) and (\ref{espurion}) we get
$$
\tau_{0i}=8 \pi b_1 {\partial u \over \partial a^i},
$$
\be
\tau_{00}=8 \pi i b_1 \Bigl(2u-\sum_i a^i{\partial u \over \partial a^i}\Bigr)
\label{taus}
\ee
Now we will present the transformation 
rules of the gauge couplings $\tau_{ij}$ under 
a monodromyy matrix $\Gamma$ in $Sp(2r, {\bf Z})$. In terms of 
the local coordinates $a_{\Gamma}^i=C^{ip}a_{D,p}(a^j,s)+ D^i_qa^q$ 
we have the couplings
\be
\tau_{\alpha \beta}^{\Gamma}=
{\partial^2 {\cal F} \over \partial a^{\alpha}_{\Gamma} 
\partial a^{\beta}_{\Gamma} }.
\label{dualcoup}
\ee
The change of coordinates is given by the matrix:
\be
\left(
      \begin{array}{cc} {\displaystyle {\partial a^i_{\Gamma} \over \partial
a^j}}& {\displaystyle {\partial
a^i_{\Gamma}
\over \partial s} }\\
& \\
{\displaystyle {\partial s \over  \partial a^j }} & 
{\displaystyle {\partial s
\over  \partial s}}
\end{array}\right) =
\left(\begin{array}{cc}{C^{ip} \tau_{pj}+D^i_j}&{C^{ip} \tau_{0p}}\\
                           0&1\end{array}\right),
\label{xv}
\ee   
with inverse
\be
\label{xvi}
\left( \begin{array}{cc} {\displaystyle{\partial a^i \over \partial
a^j_{\Gamma}}} & {\displaystyle {\partial
a^i \over \partial s }}  \\
 &  \\
     {\displaystyle {\partial s \over \partial a^j_{\Gamma}}} &
{\displaystyle {\partial s \over \partial s}}
\end{array}\right) =
 \left( \begin{array}{cc}\Big( (C\tau +D)^{-1} \Big)^i_j&
-\Big( (C\tau +D)^{-1} \Big)^i_kC^{kp}\tau_{p0}\\
                           0&1 \end{array}\right).
\ee 
Therefore we have:
$$
\Bigl( {\partial \over \partial a^j_{\Gamma}} \Big)_{\Gamma -{\rm
basis}}
=\Big( (C\tau +D)^{-1} \Big)^i_j{\partial \over \partial a^i},
$$
\be
\Big( {\partial \over \partial s} \Big)_{\Gamma -{\rm basis}}
={\partial \over \partial s}-\Big[(C\tau +D)^{-1}C \tau \Big]^i_0
{\partial \over \partial a^i};
\label{xvii}
\ee
which lead to the transformation rules for the couplings:
\bea
\tau_{ij}^{\Gamma}&=&\Big( A \tau + B \Big)\Big( C
\tau +D \Big)^{-1}_{ij},
 \,\,\,\,\,\,\,\,\,\ \tau_{0i}^{\Gamma}=
\tau_{0j}\Big( (C\tau +D)^{-1} \Big)^j_i,
\nonumber \\
\tau_{00} ^{\Gamma}&=&\tau_{00}- 
\tau_{0i}\Big[(C\tau +D)^{-1}C \tau \Big]^i_0.
\label{xviii}
\eea  
\section{Effective potential and vacuum structure}
\setcounter{equation}{0}  
In this section we will obtain, starting from the formalism 
developed in the previous section, the effective 
potential in the Coulomb phase of the softly 
broken $N=2$ theory, for a general group of 
rank $r$. 

To break $N=2$ down to $N=0$ we freeze 
the spurion superfield to a constant. 
The lowest component is 
fixed by the scale $\Lambda$, and we only 
turn on the auxiliary $F^0$ ({\it i.e.} we take 
$D^0=0$). We must include in the effective 
lagrangian $r+1$ vector multiplets, where $r$ 
is the rank of the gauge group:
\be
A^{\alpha}=(A^0, A^{I}), \,\,\,\,\,\,\ I=1, \cdots, r.
\label{twoi}
\ee
There are submanifolds in the moduli space where extra states become 
massless and we must include them in the effective lagrangian. They are 
BPS states corresponding to 
monopoles or dyons, so we introduce 
$n_H$ hypermultiplets near these submanifolds 
in the low energy description:
\be
(M_i, {\widetilde M}_i), \,\,\,\,\,\,\,\,\ i=1, \cdots, n_H
\label{twoii}
\ee
We suppose that these BPS states are mutually local, 
hence we can find a symplectic 
transformation such that they have $U(1)^r$ charges 
$(q_i^I,-q_i^I)$ with respect to the $I$-th $U(1)$ (we follow the 
$N=1$ notation). The full $N=2$ effective lagrangian contains two terms:
\be
{\cal L}={\cal L}_{\rm VM}+{\cal L}_{\rm HM}, 
\label{twoiii}
\ee
where ${\cal L}_{\rm VM}$ is given in (\ref{lagr}), and
\bea
{\cal L}_{\rm HM}&=&\sum_i\int d^4 \theta 
\big( M^{*}_{i}{\rm e}^{2q_i^IV^{(I)}}M_i 
 +{\widetilde M}^{*}_i{\rm e}^{-2q_i^IV^{(I)}}
{\widetilde M}_i\big)\nonumber\\
&+& \sum_{I,i}\Bigl( \int
d^2 \theta {\sqrt 2}A^{I}q_i^I M_i {\widetilde M}_i + {\rm h.c.} \Big)
\label{twoiv}
\eea      
The terms in (\ref{twoiii}) contributing to the effective potential are
\bea
V &=& b_{IJ}F^I {\overline F}^J + b_{0I} 
\big( F^0{\overline F}^I+{\overline F}^0F^I \big)+ 
b_{00}|F^0|^2 \nonumber\\
&+& {1 \over 2}b_{IJ} D^I D^J + 
D^I q_i^I (|m_i|^2-|{\widetilde m}_i|^2)+
 |F_{m_i}|^2+ |F_{{\widetilde m}_i}|^2 \nonumber \\
&+& \sqrt{2} \Bigl(F^I q_i^I m_i {\widetilde m}_i+
 a^Iq_i^I m_i F_{{\widetilde m}_i}+
a^Iq_i^I {\widetilde m}_i F_{m_i} +{\rm h.c.}\Bigr),
\label{twov}
\eea
where all repeated indices are summed 
and $b_{\alpha \beta}={\rm Im} \tau_{\alpha \beta} /4\pi$. 
We eliminate the auxiliary fields and obtain:
$$
D^I=-(b^{-1})^{IJ}q_i^J (|m_i|^2-|{\widetilde m}_i|^2),
$$
$$
F^I=-(b^{-1})^{IJ}b_{0J}F^0-
\sqrt{2}(b^{-1})^{IJ}q_i^J{\overline m}_i{\overline {\widetilde m}}_i,
$$
\be
F_{m_i}=
-\sqrt{2}{\overline a}^Iq_i^I{\overline {\widetilde m}}_i,
\,\,\,\,\,\,\,\ 
F_{{\widetilde m}_i}=-\sqrt{2}{\overline a}^Iq_i^I{\overline m}_i.
\label{aux}
\ee
We denote $(q_i,q_j)=\sum_{IJ}q_i^I (b^{-1})^{IJ}q_j^I$, 
$(q_i, b_0)=\sum_{IJ}q_i^I (b^{-1})^{IJ}b_{0J}$, 
$a \cdot q_i=\sum_I a^Iq_i^I$. 
Substituting in (\ref{twov}) we obtain:
\bea
V&=&{1 \over 2} \sum_{ij}(q_i,q_j)(|m_i|^2-|{\widetilde m}_i|^2)
(|m_j|^2-|{\widetilde m}_j|^2)+
2\sum_{ij}(q_i,q_j)m_i{\widetilde m}_i{\overline m}_j
{\overline {\widetilde m}}_j\nonumber\\
&+&2\sum_i|a \cdot q_i|^2(|m_i|^2+|{\widetilde m}_i|^2) +
\sqrt{2}\sum_i(q_i, b_0)\Big(F^0 m_i {\widetilde m}_i + 
{\overline F}^0{\overline m}_i{\overline {\widetilde m}}_i \Big)
\nonumber \\
&-&|F^0|^2{{\rm det}b_{\alpha \beta} \over {\rm det} b_{IJ}},
\label{twovi}
\eea
where ${\rm det} b_{\alpha \beta}/ {\rm det} b_{IJ}
=b_{00}-b_{0I}(b^{-1})^{IJ}b_{0J}$ is 
the cosmological term. This term in the 
potential is a monodromy invariant. To prove this it is 
sufficient to prove invariance under the generators 
of the symplectic group 
$Sp(2r, {\bf Z})$:
$$
\left(\matrix{A& 0 \cr
               0&(A^{\rm T})^{-1}\cr}\right), 
\,\,\,\,\,\,\,\ A\in Gl(r, {\bf Z}),
$$
\be
T_{\theta}=\left(\matrix{{\bf 1}& \theta \cr
               0&{\bf 1}\cr}\right),\,\,\,\,\,\,\,\ \theta_{ij} \in {\bf Z},
\,\,\,\,\,\,\,\,\ \Omega=\left(\matrix{0& {\bf 1} \cr
               -{\bf 1}&0\cr}\right).
\label{gener}
\ee
Invariance under $T_{\theta}$ and the matrix 
involving only $A$ is obvious, and for 
$\Omega$ one can check it easily.

The vacuum structure is determined by the minima of 
(\ref{twovi}). As in \cite{soft}, 
we first minimize with respect to $m_i$, ${\overline m}_i$:
\bea
{\partial V \over \partial{\overline m}_i}&=&
\sum_j(q_i,q_j)(|m_j|^2-|{\widetilde m}_j|^2)m_i +
2|a \cdot q_i|^2 m_i \nonumber\\
&+& 2\sum_{j}(q_i,q_j)m_j{\widetilde m}_j{\overline {\widetilde m}}_i+
\sqrt{2}{\overline F}^0 (q_i,b_0){\overline {\widetilde m}}_i=0,
\label{twovii}
\eea
\bea
{\partial V \over \partial{\widetilde {\overline m}_i} }&=&
\sum_j(q_i,q_j)(-|m_j|^2+|{\widetilde m}_j|^2){\widetilde m}_i +
2|a \cdot  q_i|^2{\widetilde m} _i \nonumber\\
&+& 2\sum_{j}(q_i,q_j)m_j{\widetilde m}_j{\overline m}_i+
\sqrt{2}{\overline F}^0 (q_i,b_0){\overline m}_i=0.
\label{twoviii}
\eea
Multiplying (\ref{twovii}) by ${\overline m}_i$, (\ref{twoviii}) by 
${\overline {\widetilde m}}_i$ and substracting, we get
\be
\sum_j(q_i,q_j)(|m_j|^2-|{\widetilde m}_j|^2)(|m_i|^2+|{\widetilde m}_i|^2) 
+
2|a \cdot q_i|^2 (|m_i|^2-|{\widetilde m}_i|^2)=0.
\label{twoix}
\ee
Suppose now that, for some indices 
$i \in I$, $|m_i|^2+|{\widetilde m}_i|^2 >0$. 
Multiplying (\ref{twoix}) by $|m_i|^2-|{\widetilde m}_i|^2$ 
and summing over $i$ we obtain
\be
\sum_{ij}(q_i,q_j)(|m_i|^2-|{\widetilde m}_i|^2)
(|m_j|^2-|{\widetilde m}_j|^2)=
-\sum_{i \in I}{2|a \cdot  q_i|^2 \over |m_i|^2+|{\widetilde m}_i|^2}
(|m_i|^2-|{\widetilde m}_i|^2)^2.
\label{twox}
\ee
The matrix $(b^{-1})^{IJ}$ is positive definite, 
and if the charge vectors $q_i^I$ are 
linearly independent it 
follows that the matrix $(q_i, q_j)$ is positive 
definite too. Then the l.h.s. of 
(\ref{twox}) is $\ge 0$ while the r.h.s. is $\le 0$. 
The only way for this equation to be 
consistent is if
\be
|m_i|=|{\widetilde m}_i|, \,\,\,\,\,\,\ i=1, \cdots, n_H.
\label{twoxi}
\ee
In this case we can write the equation (\ref{twovii}), after 
absorbing the phase of $F^0=f_0{\rm e}^{i\gamma}$ in ${\widetilde m}_i$, as:
\be
2|a \cdot q_i|^2 m_i + 
2\sum_{j}(q_i,q_j)m_j{\widetilde m}_j{\overline {\widetilde m}}_i+
\sqrt{2}f_0 (q_i,b_0){\overline {\widetilde m}}_i=0.
\label{twoxii}
\ee
Multiplying by ${\overline m}_i$ and summing over $i$, we obtain
\be
2\sum_i |a \cdot q_i|^2 |m_i|^2 +
\sqrt{2}f_0 \sum_i(q_i,b_0){\overline m}_i{\overline {\widetilde m}}_i=
-2\sum_{ij}(q_i,q_j)m_j{\overline m}_i{\widetilde m}_j
{\overline {\widetilde m}}_i,
\label{twoxiii}
\ee
hence $\sqrt{2}f_0 
\sum_i(q_i,b_0){\overline m}_i{\overline {\widetilde m}}_i$ is real. We 
can insert \label{twoxiii} in (\ref{twovi}) and get 
the following expression for the effective potential:
\be
V=-f_0^2{{\rm det}b_{\alpha \beta} \over {\rm det} b_{IJ}}-
2\sum_{ij}(q_i,q_j)m_j{\overline m}_i{\widetilde m}_j
{\overline {\widetilde m}}_i.
\label{twoxiv}
\ee
If (\ref{twoxi}) holds, we can fix the gauge 
in the $U(1)^r$ factors and write
\be
m_i=\rho_i, \,\,\,\,\,\,\ {\widetilde m}_i=\rho_i {\rm e}^{i\phi_i}
\label{twoxv}
\ee
and (\ref{twoxii}) reads:
\be
\rho_i^2 \Bigl(|a\cdot q_i|^2+
\sum_{j}(q_i,q_j)\rho^2_j{\rm e}^{i(\phi_j-\phi_i)}+
{f_0 (q_i,b_0) \over \sqrt{2}}{\rm e}^{-i\phi_i}\Bigr)=0.
\label{twoxvi}
\ee
Apart form the trivial solution $\rho_i=0$, we have:
\be
|a\cdot q_i|^2+\sum_{j}(q_i,q_j)\rho^2_j{\rm e}^{i(\phi_j-\phi_i)}+
{f_0 (q_i,b_0) \over \sqrt{2}}{\rm e}^{-i\phi_i}=0
\label{twoxvii}
\ee
and we can have a monopole (or dyon) VEV in some 
regions of the moduli space. Notice that 
for groups of rank $r>1$ there is a coupling 
between the different $U(1)$ factors and one
 needs a numerical study of the equation 
above once the values of the charges $q_i^I$ are known. 
In addition, the moduli 
space is in that case very complicated and explicit
 solutions for the prepotential and
 gauge couplings of the $N=2$ theory are
 difficult to find. However we still can have 
some qualitative information in many cases
 under some mild assumptions, as we will see.

\section{Vacuum structure of the $SU(N)$ Yang-Mills theory}
\setcounter{equation}{0}
The moduli space of vacua of the $N=2$ $SU(N)$ Yang-Mills can be 
parametrized in a gauge-invariant way by the elementary symmetric polynomials 
$s_{l}$, $l=2, \cdots, N$ in the 
eigenvalues of $\langle \phi \rangle$, $\phi_i$. The vacuum 
structure of the theory is associated to 
the hyperelliptic curve \cite{kl}:
$$
y^2=P(x)^2-{\Lambda}^{2N},
$$
\be
P(x)={1 \over 2}{\rm det}(x-\langle \phi \rangle)=
{1 \over 2}\prod_i(x-\phi_i),
\label{fouri}
\ee
where $\Lambda$ is the dynamical scale of the $SU(N)$ theory and $P(x)$
 can be written in terms of the variables 
$s_l$ as $P(x)={1/2}\sum_l(-1)^ls_lx^{N-l}$. 
Once the hyperelliptic curve is known, one can 
compute in principle the metric 
on the moduli space and the exact quantum prepotential,
 but explicit solutions are 
difficult to find (they have been obtained in \cite{klt} for the $SU(3)$ case). 
However, as in the $SU(2)$ case \cite{soft}, one expects that the minima
 of the effective potential for 
the $SU(N)$ theory are near the $N=1$ points 
(at least for a small supersymmetry 
breaking parameter). The physics of the $N=1$ points 
in $SU(N)$ theories has a much 
simpler description because it involves only small 
regions of the moduli space, and 
has been studied in \cite{ds}. The $N=1$ points correspond 
to points in the moduli 
space where $N-1$ monopoles or dyons coupling to each $U(1)$ 
become massless simultaneously. 
From the point of view of the hyperelliptic curve, 
the $N=1$ point where $N-1$ monopoles become massless 
($a_{D,I}=0$) corresponds to a 
simultaneous degeneration 
of the $N-1$ $\alpha$-cycles, associated to magnetic monopoles. 
This means in turn that 
the polynomial $P(x)^2-\Lambda^{2N}$ must have $N-1$ double zeros and two 
single zeros. If we set $\Lambda=1$, this can be achieved with the Chebyshev 
polynomials
\be
P(x)={\rm cos}\Big(N{\rm arccos} {x\over 2}\Big),
\label{fourii}
\ee
and the corresponding eigenvalues are 
$\phi_i =2 {\rm cos} \pi(i-{1 \over 2})/N$. 
The other 
$N-1$ points, corresponding to the simultaneous condensation of $N-1$ 
mutually local dyons, are obtained with the action of the anomaly-free 
discrete subgroup 
${\bf Z}_{4N} \subset U(1)_R$. One can perturb slightly the curve 
(\ref{fourii}) to obtain the 
effective lagrangian (or equivalently, 
the prepotential) at lowest order. What is found 
is that, in terms of the dual monopole variables 
$a_{D,I}$, the $U(1)$ factors are 
decoupled and $\tau^D_{IJ} \sim \delta_{IJ}\tau_I$. 
Near the $N=1$ point where $N-1$ 
monopoles 
become massless one can then simplify the equation 
(\ref{twoxvii}) for the monopole 
VEVs, because $q_i^I=\delta_i^I$, 
$(b^{-1})^{IJ}=\delta^{IJ} b^{-1}_I$. The equation reduces 
then to $r=N-1$ $SU(2)$-like equations, 
and in particular the phase factors 
${\rm e}^{-i\phi_I}$ must be real. We then 
set ${\rm e}^{-i\phi_I}=\epsilon_I$, 
$\epsilon_I=\pm 1$. The VEVs are determined by:
\be
 \rho^2_I=-b_I|a_{D,I}|^2-
{f_0 b_{0I}\epsilon_I \over \sqrt{2}},\,\,\,\,\,\,\ I=1, \cdots, r.
\label{fouriii}
\ee
The effective potential (\ref{twoxiv}) reads:
\be
V=-f_0^2 \Big(b_{00}-\sum_I{b_{0I}^2 \over b_I} \Big)
-2\sum_I{1 \over b_I}\rho_I^4.
\label{fouriv}
\ee
The quantities that control, at least qualitatively, 
the vacuum structure of the theory, 
are $b_{0I}$ and $b_{00}$. If $b_{0I} \not= 0$ 
at the $N=1$ points, we have a monopole VEV 
for $\rho_I$ around this point. If $b_{0I}= 0$, we still 
can have a VEV, as it happens in 
the $SU(2)$ case in the dyon region, but we expect that 
it will be too tiny to produce a local minimum \cite{soft}. When 
one has monopole or dyon condensation at one of these $N=1$ 
points in all the $U(1)$ factors, the 
value of the potential at this point is given by
\be
V=-f_0^2b_{00},
\label{fourv}
\ee
and if the local minimum is very near to the $N=1$ point,
 we can compare the energy of the 
different $N=1$ points according to (\ref{fourv}) 
and determine the true vacuum of the 
theory. Hence, to have a qualitative picture
 of the vacuum structure, and if we 
suppose that the minima of the effective potential
 will be located near the $N=1$ points, 
we only need to evaluate $b_{0I}$, $b_{00}$ at
 these points. This can be done using the 
explicit solution in \cite{ds} and the expressions (\ref{taus}). 

To obtain the correct normalization of the constant appearing in 
(\ref{espurion}) we can evaluate $\sum_{I}(a_{D,I}da/du-ada_{D,I}/du)$ 
in the $N=1$ points, obtaining the constant value $4 \pi i b_1$. 
 The value of the quadratic Casimir at the $N=1$ point described by 
(\ref{fourii}) is
\be
u=\langle {\rm Tr}\phi^2 \rangle =
4\sum_{i=1}^N {\rm cos}^2 {\pi(i-1/2) \over N} =2N,
\label{fourvi}
\ee
and the values at the other $N=1$ 
points are given by the action of ${\bf Z}_N$ ($u$ has 
charge $4$ under $U(1)_R$):
 $u^{(k)}=2\omega^{4k}N$, $\omega={\rm e}^{\pi i /2N}$ with 
$k=0, \cdots, N-1$. To compute $\tau_{0I}$ we must 
also compute $\partial u /\partial a_{D,I}$. 
Using the results of \cite{ds}, we have:
\be
{\partial u \over \partial a_{D,I}}=-4i {\rm sin}{\pi I \over N},
\label{fourvii}
\ee
and using $b_1=2N/16 \pi^2$, we obtain
\be
\tau_{0I}=4\pi b_1{\partial u \over \partial a_{DI}}=
 -{2Ni \over \pi}{\rm sin}{\pi I \over N}.
\label{fourviii}
\ee
At the $N=1$ point where $N-1$ monopoles condense, $a_{D,I}=0$, therefore
\be
\tau_{00}=8\pi i u ={2i \over \pi}N^2.
\label{fourix}
\ee
(\ref{fourviii}) indicates that 
monopoles condense at this point in all the 
$U(1)$ factors, but with different VEVs. 
This is a consequence of the spontaneous 
breaking of the $S_N$ symmetry permuting the $U(1)$ factors \cite{ds}. 

To study the other $N=1$ points we must 
implement the ${\bf Z}_N$ symmetry in the 
$u$-plane. The local coordinates $a^{(k)}_I$
 vanishing at these points are given by 
a $Sp(2r, {\bf Z})$ transformation acting on 
the coordinates $a_I$, $a_{D,I}$ 
around the monopole point. The ${\bf Z}_N$ symmetry implies that
\be
{\partial u \over \partial a^{(k)}_{I}}(u^{(k)})=
\omega^{2k}{\partial u \over \partial a_{D,I}}(u^{(0)}),
\label{fourx}
\ee
and then we get
$$
b^{(k)}_{0I}={1 \over 4 \pi}{\rm Im}\tau^{(k)}_{0I}=
-{N \over 2\pi^2}{\rm cos}{\pi k \over N}{\rm sin}{\pi I \over N},
$$
\be
b^{(k)}_{00}=
{1 \over 4 \pi}{\rm Im}\tau^{(k)}_{00}=
{1 \over 2}\Bigl({N \over \pi} \Bigr)^2
{\rm cos}{2\pi k \over N}.
\label{fourxi}
\ee
The first equation tells us that generically we will have dyon 
condensation at all the $N=1$
points, and the second equation together with (\ref{fourv}) implies 
that the condensate 
of $N-1$ monopoles at $u=2N$ is energetically favoured, 
and then it will be the true 
vacuum of the theory. Notice that the ${\bf Z}_N$ 
symmetry works in 
such a way that the size of the condensate, given by 
$|{\rm cos}{\pi k \over N}|$, corresponds to an energy given by 
$-{\rm cos}{2\pi k \over N}$: as one should expect, the bigger the 
condensate the smaller its energy. In fact, for 
$N$ even the $N=1$ point corresponding to $k=N/2$ has 
no condensation. In this case the energy is still 
given by (\ref{fourv}), as the effective potential equals 
the cosmological term with $b_{0I}=0$, 
and is the biggest one.

\section{Mass formula in softly broken $N=2$ theories}
\subsection{A general mass formula}
In some cases the mass spectrum of a 
softly broken supersymmetric theory is such that the 
graded trace of the square of the mass matrix is zero as it 
happens in supersymmetric theories 
\cite{mass}. We will see in this section that 
this is also the case when we softly break 
$N=2$ supersymmetry with a dilaton spurion. 

We will then compute the trace of the squared mass 
matrix which arises from the effective lagrangian 
(\ref{twoiii}), once the supersymmetry breaking parameter is turned on. 
The fermionic content of the theory is 
as follows: we have fermions $\psi^I$, $\lambda^I$ 
coming from the $N=2$ vector multiplet $A^I$ 
(in $N=1$ language, $\psi^I$ comes from the $N=1$ 
chiral multiplet and $\lambda^I$ from the $N=1$ vector multiplet). 
We also have 
``monopolinos" $\psi_{m_i}$, $\psi_{{\widetilde m}_i}$ 
from the $n_H$ matter hypermultiplets. 
To obtain the fermion mass matrix, we just
 look for fermion bilinears in (\ref{twoiii}). From 
the gauge kinetic part and the K\"ahler potential in ${\cal L}_{\rm VM}$
 we obtain:
\be
{i \over 16 \pi} F^{\alpha} {\partial}_{\alpha}\tau_{IJ}\lambda^I \lambda^J +
{i \over 16 \pi} {\overline F}^{\alpha} 
{\partial}_{\alpha}\tau_{IJ}\psi^I \psi^J.
\label{fivei}
\ee
where $F^0=f_0$ and the auxiliary fields $F^I$ are given in (\ref{aux}).
 From the kinetic term and the superpotential in ${\cal L}_{\rm HM}$ we get:
\bea
& & i\sqrt{2}\sum_{i} q_i \cdot \lambda \bigl({\overline m}_i\psi_{m_i} - 
{\overline{\widetilde m}}_i \psi_{{\widetilde m}_i} \bigr) \nonumber\\
&-& \sqrt{2} \sum_i \Bigl(a \cdot q_i\psi_{m_i}\psi_{{\widetilde m}_i}+
q_i \cdot \psi \psi_{{\widetilde m}_i} m_i +
q_i \cdot \psi \psi_{m_i} {\widetilde m}_i \Bigr)
\label{fiveii}
\eea
If we order the fermions as $(\lambda, \psi, 
\psi_{m_i}, \psi_{{\widetilde m}_i})$ and
denote $\mu^{IJ}=iF^{\alpha} {\partial}_{\alpha}\tau_{IJ} /4\pi$, 
${\hat \mu}^{IJ}=i{\overline F}^{\alpha} {\partial}_{\alpha}\tau_{IJ}/4\pi$, 
the ``bare" fermionic mass matrix reads:
\be
M_{1/2}= \left( \begin{array}{cccc} \mu /2& 0 & i\sqrt{2} q_i^I{\overline m}_i
&-i\sqrt{2} q_i^I{\overline {\widetilde m}}_i \\
0& {\hat \mu}/2& -\sqrt{2} q_i^I{\widetilde m}_i& -\sqrt{2} q_i^I m_i\\
i\sqrt{2} q_i^I{\overline m}_i&-
\sqrt{2} q_i^I{\widetilde m}_i&0&-\sqrt{2}a \cdot q_i\\
-i\sqrt{2} q_i^I{\overline {\widetilde m}}_i&
-\sqrt{2} q_i^I m_i&-\sqrt{2}a \cdot q_i& 0 \end{array}\right),
\label{fiveiii}
\ee
but we must take into account the wave function renormalization 
for the fermions $\lambda^I$, 
$\psi^I$ and consider
\be
{\cal M}_{1/2} =Z M_{1/2} Z, \,\,\,\,\,\,\,\ 
Z=\left( \begin{array}{cccc} b^{-1/2} & 0 & 0
&0 \\
0& b^{-1/2}& 0&0\\
0&0&1&0\\
0&
0&0& 1 \end{array}\right)
\label{fiveiv}
\ee
The trace of the squared fermionic matrix can be easily computed:
\bea
{\rm Tr}{\cal M}_{1/2} {\cal M}^{\dagger}_{1/2}&=&
{1\over 4}{\rm Tr}[\mu b^{-1} {\overline \mu}b^{-1}+
{\hat \mu}b^{-1}{\overline {\hat \mu}} 
b^{-1}]\nonumber\\
&+& 4\sum_i |a \cdot q_i|^2 + 
8\sum_i(q_i, q_i)(|m_i|^2+|{\widetilde m}_i|^2).
\label{fivev}
\eea
The scalars in the model are the monopole fields 
$m_i$, ${\widetilde m}_i$ and 
the lowest components of the $N=1$ chiral 
superfields in the $A^I$, $a^I$. To compute the
 trace of the scalar mass matrix we need
$$
{\partial^2 V \over \partial m_i \partial {\overline m}_i}= 
\sum_l (q_i, q_l) 
(|m_l|^2-|{\widetilde m}_l|^2)+ 
(q_i, q_i)(|m_i|^2+2|{\widetilde m}_i|^2)+2|a \cdot q_i|^2,
$$
$$
{\partial^2 V \over \partial {\widetilde m}_i \partial
{\overline {\widetilde m}_i} }= -\sum_l (q_i, q_l) 
(|m_l|^2-|{\widetilde m}_l|^2)+ 
(q_i, q_i)(2|m_i|^2+|{\widetilde m}_i|^2)+2|a \cdot q_i|^2,
$$
\bea
{\partial^2 V \over \partial a^I \partial {\overline a}^J}
&=&f_0^2{\partial^2 (b_0,b_0) \over \partial a^I \partial {\overline a}^J}+
2\sum_{k,l}{\partial^2 (q_k,q_l) \over \partial a^I \partial {\overline a}^J} 
m_k{\widetilde m}_k {\overline m}_l {\overline {\widetilde m}}_l\nonumber\\
&+&2\sum_{k}q_k^Iq_k^J(|m_k|^2+|{\widetilde m}_k|^2)\nonumber\\
&+&
\sqrt{2}\sum_k{\partial^2 (q_k,b_0)  
\over \partial a^I \partial {\overline a}^J}f_0
(m_k{\widetilde m}_k +{\overline m}_k{\overline{\widetilde m}}_k).
\label{fivevi}
\eea
In the last expression we used that, due to the holomorphy of the 
couplings $\tau_{\alpha \beta}$, 
$\partial^2_{I {\overline J}}b_{\alpha \beta}=0$.
 If we assume that we are in the conditions of section 2, 
at the minimum we have $|m_i|=
|{\widetilde m}_i|$, and the trace of the squared scalar matrix is
\be
{\rm Tr}{\cal M}^2_{0}=
 6\sum_i (q_i,q_i)(|m_i|^2+|{\widetilde m}_i|^2)+8\sum_i |a \cdot q_i|^2+ 
2(b^{-1})^{IJ}{\partial^2 V \over \partial a^I \partial {\overline a}^J},
\label{fivevii}
\ee
where we have included the wave function renormalization for the 
scalars $a^I$. The mass of the dual photon is given by the monopole 
VEV through the magnetic Higgs 
mechanism:
\be
{\rm Tr}{\cal M}^2_1=2\sum_i(q_i, q_i) ( |m_i|^2+|{\widetilde m}_i|^2).
\label{fiveviii}
\ee
Taking into account all these contributions, 
the graded trace of the squared matrix is:
\bea
& &\sum_j (-1)^{2j}(2j+1){\rm Tr }{\cal M}^2_j=
-{1\over 2}{\rm Tr}[\mu b^{-1} {\overline \mu}b^{-1}
+{\hat \mu}b^{-1}{\overline {\hat \mu}} 
b^{-1}] \nonumber\\
& &\,\,\,\,\,\,\,\ +
2f_0^2{\rm Tr}b^{-1}\partial {\overline \partial}(b_0,b_0)+
4\sum_{k,l}{\rm Tr}b^{-1}\partial {\overline \partial}(q_k,q_l)
m_k{\widetilde m}_k {\overline m}_l {\overline {\widetilde m}}_l\nonumber\\
& &\,\,\,\,\,\,\,\ +
2\sqrt{2}\sum_k{\rm Tr}b^{-1}\partial {\overline \partial}(q_k,b_0) f_0
(m_k{\widetilde m}_k +{\overline m}_k{\overline{\widetilde m}}_k).
\label{traza}
\eea
To see that this is zero, we write the bilinears in the monopole 
fields in terms of the auxiliary fields 
$F^I$, ${\overline F}^I$, using (\ref{aux}):
\be
\sum_{i} q_i^I{\overline m}_i{\overline {\widetilde m}}_i=
-{1 \over \sqrt{2}}(b_{IJ}F^J+b_{0I}f_0).
\label{fiveix}
\ee
Then we can group the terms in (\ref{traza}) depending on the number of 
$F^I$, ${\overline F}^I$, and check 
that they cancel separately. For instance, for the 
terms with two auxiliaries, we have from the first term in (\ref{traza}):
\be
-2(F^I {\overline F}^J +{\overline F}^I F^J)\partial_I b_{MN}(b^{-1})^{NP}
\partial_{\overline J}b_{PQ}(b^{-1})^{QM}
\label{fivex}
\ee
and from the third term
\bea
& &2F^I {\overline F}^J\partial_M b_{JN}(b^{-1})^{NP}
\partial_{\overline Q}b_{PI}(b^{-1})^{QM}\nonumber\\
&+&2F^I {\overline F}^J\partial_M b_{PI}(b^{-1})^{NP}
\partial_{\overline Q}b_{JN}(b^{-1})^{QM}.
\label{fivexi}
\eea
Taking into account the holomorphy of the couplings and the K\"ahler 
geometry, we have 
$\partial_M b_{PI}=\partial_I b_{PM}$, 
$\partial_{\overline Q}b_{JN}=
\partial_{\overline J}b_{QN}$, so (\ref{fivex}) and 
(\ref{fivexi}) add up to zero. With a little more algebra one can verify 
that the terms with one $F^I$ (and their conjugates with ${\overline F}^I$)
 and without any auxiliaries add up to zero too. The result is then:
\be
\sum_j (-1)^{2j}(2j+1){\rm Tr }{\cal M}^2_j=0.
\label{fivexii}
\ee

\subsection{Mass spectrum in the $SU(2)$ case}
In the $SU(2)$ case we can obtain much more information about the mass matrix 
and also determine its eigenvalues. First we consider the fermion mass
matrix. Taking into account that at the minimum of the effective potential 
$m = {\overline m}=\rho$, ${\tilde m}=\epsilon m$, we can introduce the linear 
combination:
\be
\eta_{\pm}={1\over {\sqrt 2}} (\psi_m \pm \epsilon \psi_{\tilde m}).
\label{fivexiii}
\ee
With respect to the new fermion fields $(\lambda, \eta_{+}, \psi,\eta_{-})$, 
the bare fermion mass matrix reads:
\be
M_{1/2}= \left( \begin{array}{cccc} {1 \over 2} \mu & -2\epsilon \rho & 0
&0\\
-2\epsilon \rho&-{\sqrt 2}\epsilon a & 0& 0\\
0&0&{1 \over 2} \mu &2i\rho\\
0&0&2i\rho & -{\sqrt 2}\epsilon a  \end{array}\right),
\label{fivexiv}
\ee
Notice that, in the $SU(2)$ case, the auxiliary field $F$ is real 
and $\mu={\hat \mu}$. ${\cal M}_{1/2} {\cal M}^{\dagger}_{1/2}$ can be 
easily diagonalized. From (\ref{fivexiv}) it is easy to see
that the squared fermion mass matrix is block-diagonal with the 
same $2\times 2$ matrix in both entries:
\be
\left(\begin{array}{cc} b_{11}^{-2}\mu {\overline \mu}/4 
+4b_{11}^{-1}\rho^2& -\epsilon b_{11}^{-3/2}\mu\rho+
 2{\sqrt 2}{\overline a}\rho \\
-\epsilon b_{11}^{-3/2}{\overline \mu}\rho+
 2{\sqrt 2} a\rho &4b_{11}^{-1}\rho^2+2|a|^2\end{array} \right).
\label{fermimat}
\ee
Hence there are two different doubly degenerate eigenvalues. In terms of 
the determinant and trace of (\ref{fermimat}),
\bea
\alpha &=& (m^{F}_1)^2+(m^{F}_2)^2=
{1\over 4b_{11}^2}\mu{\overline \mu}+
2|a|^2+{8\over b_{11}}\rho^2\nonumber\\
\beta &=&(m^{F}_1)^2(m^{F}_2)^2 ={1\over b_{11}^2}|4\rho^2 + {\epsilon \over 
{\sqrt 2}}a\mu|^2,
\label{fivexv}
\eea
the eigenvalues are:
\be
(m^{F}_{1,2})^2={\alpha \over 2}\pm {1 \over 2}{\sqrt {\alpha^2-4\beta}}.
\label{fivexvi}
\ee
The computation of the scalar mass matrix is more lengthy. First we must 
compute the second derivatives of the effective potential, evaluated at the 
minimum. To obtain more simple expressions, we can use the identities 
(\ref{taus}) to express all the derivatives of the couplings in terms 
only of $\partial b_{11}/\partial a$, $\partial^2 b_{11}/\partial a^2$.
The results are:
\bea
{\partial^2 V \over \partial m \partial {\overline m}}&=&{3\over b_{11}}\rho^2
+2|a|^2, \,\,\,\,\,\ {\partial^2 V/\partial m^2}=
{\partial^2 V \over \partial {\widetilde m}^2}={1\over b_{11}}\rho^2, 
\,\,\,\,\,\ \nonumber \\ 
{\partial^2 V \over \partial m \partial {\widetilde m}}&=&
{\epsilon\over b_{11}}\rho^2+ {{\sqrt 2} b_{01} \over b_{11}}f_0, 
\,\,\,\,\,\ 
{\partial^2 V\over \partial m \partial {\overline{\widetilde m}}}=
{\epsilon\over b_{11}}\rho^2 \nonumber\\  
{\partial^2 V \over \partial m \partial a}&=& 
2\rho\Big[{\overline a}-
\big(b_{11} {\partial \over \partial a}{1 \over b_{11}}
\big) \big(|a|^2-{i \epsilon \over {\sqrt 2}}af_0 \big)\Big]\nonumber\\
{\partial^2 V \over \partial a^2}&=&
-b_{11}^2 \big({\partial \over \partial a}{1 \over b_{11}}\big) f_0
(af_0+2 {\sqrt 2} i\epsilon |a|^2)\nonumber \\
& & -b_{11}^2 \big({\partial^2 \over \partial a^2}{1 \over b_{11}}\big)
(af_0+{\sqrt 2}i\epsilon |a|^2)^2,\nonumber\\
{\partial^2 V \over \partial {\widetilde m} \partial a}&=&
\epsilon {\partial^2 V \over \partial m \partial a},\,\,\,\,\,\ 
{\partial^2 V \over \partial {\overline m} \partial a}=
 {\partial^2 V \over \partial m \partial a},\,\,\,\,\,\
{\partial^2 V \over \partial {\overline {\widetilde m}} \partial a}=
{\partial^2 V \over \partial {\widetilde m} \partial a},\nonumber\\
{\partial^2 V \over \partial {\overline a} \partial a}&=&
4\rho^2+{1 \over 2b_{11}}\mu {\overline \mu},
\label{fivexvii}
\eea
and the rest of the derivatives are obtained through complex 
conjugation. In the last line 
we used the result of the previous section.
 To obtain the bosonic mass matrix we must take into account the wave-function 
renormalization of the $a$, ${\overline a}$ 
variables, as in (\ref{fivevii}).  
Its eigenvalues are as follows: we have a zero 
eigenvalue corresponding to 
the Goldstone boson of the spontaneously broken 
$U(1)$ symmetry. There is 
also an eigenvalue with degeneracy two given by:
\be
2\big({\partial^2 V \over \partial m \partial {\overline m}}-
{\partial^2 V \over \partial {\widetilde m}^2}\big)= 
-{2 {\sqrt 2} \epsilon \over b_{11}}f_0b_{01}.
\label{fivexviii}
\ee
Notice that this is always positive if we have a non-zero VEV for $\rho$. 
The other three eigenvalues are best 
obtained numerically, as they are the solutions 
to a  
third-degree algebraic equation. 

As an application of these general results, 
we can plot the mass spectrum 
as a function of the supersymmetry breaking 
parameter $f_0$ in the $SU(2)$ Yang-Mills case, 
where the minimum corresponds to the monopole region 
and $\epsilon=-1$. We have only to 
compute the derivatives of the magnetic coupling. Using 
the elliptic function representation of the 
Seiberg-Witten solution \cite{soft} we obtain:
\be
{\partial \tau^{(m)}_{11} \over \partial a^{(m)}}={\pi^2 \over 8}{k \over 
k'^2 K'^3},\,\,\,\,\,\
{\partial^2 \tau^{(m)}_{11} \over \partial {a^{(m)}}^2}=
-{\pi i \over 32} {k^2 \over k'^4 K'^4}\Big( k'^2-k^2+{3E' \over K'}\Big),
\label{fivexix}
\ee
where we set $\Lambda=1$. 
\figalign{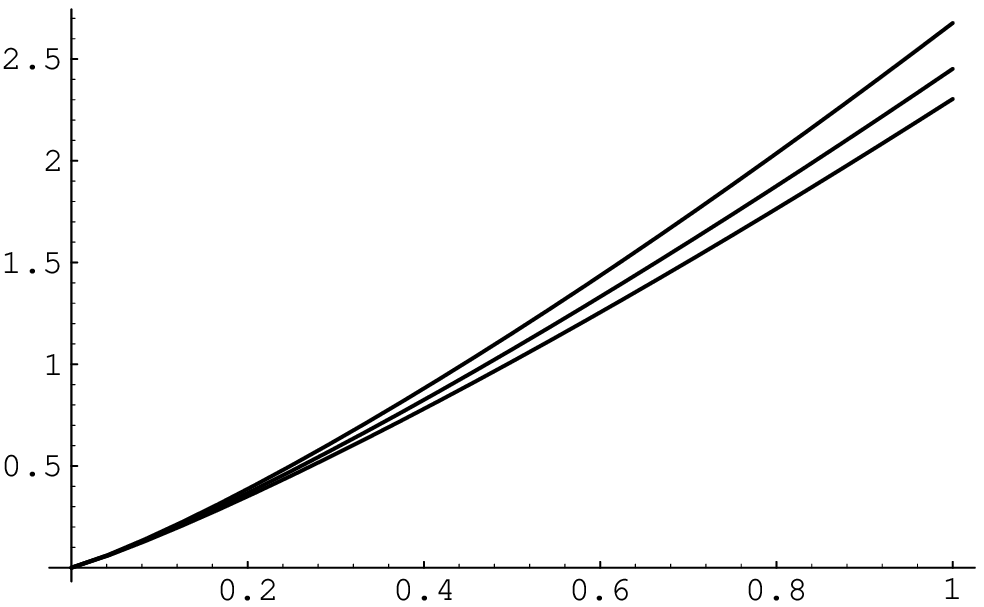}{\fer}{Fermion masses (\ref{fivexvi})
(top and bottom) 
and photon mass (\ref{fiveviii}) (middle) in softly broken 
$SU(2)$ Yang-Mills, 
as a function of $0 \le f_0 \le 1$.}{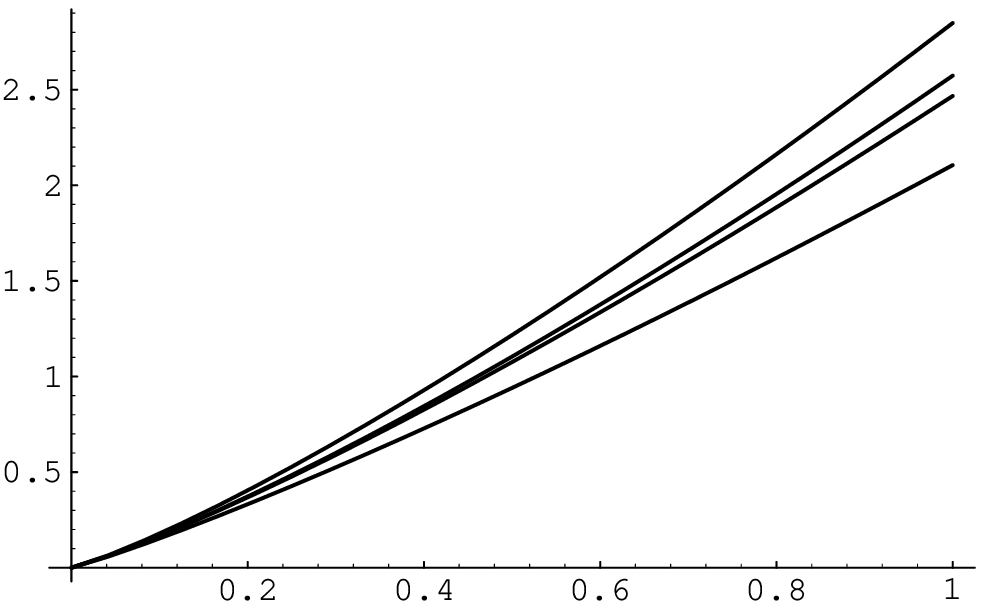}{\sca}{Masses of the
 scalars in softly broken 
$SU(2)$ Yang-Mills, as a function of $0 \le f_0\le 1$.}

These derivatives diverge at the monopole singularity 
$u=1$, and we may think that this can give some kind of singular behaviour 
for the masses near this point. In fact this is not so. The position of 
the minimum, $u_0$, behaves almost 
linearly with respect to $f_0$, $u_0-1 \sim f_0$, and this guarantees that 
the behaviour very near to $u=1$ (corresponding to a very small $f_0$) is 
perfectly smooth, as one can see in the figures. In \fig{\fer} we plot 
the fermion masses (\ref{fivexvi}) (top and bottom) and the photon mass 
given in (\ref{fiveviii}) (middle). In \fig{\sca} we plot the masses of 
the scalars, where the second one from the top corresponds to 
the doubly degenerate eigenvalue (\ref{fivexviii}).

\section{$SU(2)$ theory with one massless hypermultiplet}
\setcounter{equation}{0}

The low energy effective action of $N=2$ 
supersymmetric QCD with one massless hypermultiplet is described 
by the elliptic curve \cite{swtwo}:
\be
y^2 =x^3-ux^2-{1\over 64} \Lambda_1^6
\label{sixi}
\ee
with discriminant 
\be
\Delta=-{\Lambda_1^6 \over 16}[u^3+ {27\over 256}\Lambda_1^6]
\label{sixii}
\ee
and there are three singularities in the $u$-plane corresponding to the 
roots of the equation $u^3=-27\Lambda_1^6/256$. These singularities are 
related by a ${\bf Z}_3$ symmetry coming from the anomaly-free 
discrete subgroup ${\bf Z}_{12} 
\subset U(1)_R$. In what follows we will normalize the 
dynamical scale of the theory as $\Lambda_1^6= 256/27$, so the 
three singularities of the $u$-plane are located at the 
three roots $u_1={\rm e}^{-i\pi/3}$, $u_2={\rm e}^{i\pi/3}$ 
and $u_3=-1$. These singularities correspond to massless 
BPS states with quantum numbers $(n_e, n_m)=(0,1)$, $(1,1)$ and 
$(2,1)$. The periods of the curve (\ref{sixi}) $a(u)$, $a_D(u)$ 
satisfy the Picard-Fuchs equation \cite{iy}:
\be
{d^2 \omega \over du^2}+{1\over 4}{u \over u^3+1}\omega=0,
\label{sixiii}
\ee
and as in the $N_f=0$ case the wronskian $ada_D/du-a_Dda/du$ is a constant 
(this is in fact the case for all the $SU(2)$ theories with $N_f \le 3$). 
This constant appears in the r.h.s. of (\ref{espurion}) and depends on the 
normalization chosen for the periods. The normalization of \cite{swtwo} is 
such that the electric charges are integers, and then the periods behave 
near infinity as
\bea
a(u) &\sim & {1\over 2}{\sqrt {2u}} \nonumber\\
a_D(u)  &\sim & i {4-N_f \over 2 \pi}a(u){\rm ln}{ u \over \Lambda_{f}^2}.
\label{sixiv}
\eea
This gives for the wronskian the constant value $i(4-N_f )/4 \pi$, hence 
in the r.h.s. of (\ref{espurion}) and with this normalization we must use 
$4\pi b_1$, where $b_1=(4-N_f )/16 \pi^2$ is the coefficient of the 
one-loop $\beta$-function.  

The solution to (\ref{sixiii}) has been obtained in \cite{iy} and further 
ellaborated in \cite{bf}, where the global issues are carefully analyzed. It is 
expressed in terms of hypergeometric functions and the branch cuts must be 
taken into account. For $a(u)$ we have:
\be
a(u)= {1\over 2}{\sqrt {2u}} F(-{1\over 6},{1\over 6},2;-{1 \over u^3}),
\label{sixv}
\ee
with one branch cut along the negative real axis corresponding to the 
square root, and three branch cuts going from the three singularities to 
the origin and corresponding to the branch cut of the hypergeometric function. 
To obtain the solution for $a_D(u)$ we must choose at what singularity 
the monopole $(0,1)$ becomes massless, and then the asymptotic behaviour 
\label{sixiv} fixes the solution. Following \cite{bf}, we choose $u_1={\rm 
e}^{-i \pi/3}$ as the monopole singularity. If we now define the function    
\be
f_D(u)={{\sqrt {12}}\over 12}(u^3+1)F({5\over 6},{5\over 6},2;1+u^3),
\label{sixvi}
\ee
the other period $a_D(u)$ will be given by the analytic continuation 
of $f_D(u)$ through the branch cuts of the hypergeometric function in 
(\ref{sixvi}). Explicitly:
\bea
-{2\pi \over 3}< {\rm Arg}u <0 &:& a_D(u)={\rm e}^{-2\pi i/3}f_D(u),
\nonumber\\
0< {\rm Arg}u < {2\pi \over 3} &:& a_D(u)={\rm e}^{-2\pi i/3}f_D(u)-a(u),
\nonumber\\
{2\pi \over 3}< {\rm Arg}u < \pi  &:& a_D(u)=f_D(u)-2a(u),
\nonumber\\
-\pi< {\rm Arg}u <-{2\pi \over 3}   &:& a_D(u)=-f_D(u)+a(u).
\label{sixvii}
\eea
As $f_D(u)=0$ vanishes at the three singularities, we see that 
the good variable around $u_2$ is $a_D(u)+a(u)$, and thus corresponds to the 
dyon $(1,1)$. For the dyon becoming massless at $u_3=-1$ there is a 
branch cut and two different descriptions, one with quantum numbers 
$(2,1)$ in the upper half of the complex $u$-plane (and the corresponding 
coordinate is $a_D(u)+2a(u)$), and another one with 
quantum numbers $(-1,1)$ in 
the lower half, corresponding to the coordinate $a_D(u)-a(u)$. 
We then see that 
$a_D(u)$ has two branch cuts, one along the negative real axis, and another 
one going from $u_2$ to the origin. If we recall that 
$a_D(u)$ is the good coordinate to describe the $(0,1)$ monopole, 
we can see that these cuts are due to the singularities 
associated with BPS states which are non-local with respect to 
the monopole. Another interesting feature of (\ref{sixvii}) is the explicit 
realization of the ${\bf Z}_3$ symmetry acting on the $u$-plane.  

In the softly broken theory $N_f=1$ with a dilaton spurion we must compute the 
couplings (\ref{dilcouplings}) using (\ref{taus}) with the right normalization. 
All we need are the explicit expressions for 
$da/du$, $df_D/du$. Using the properties of hypergeometric functions, 
one obtains:
\bea
{da \over du}&=&{ 1\over 2 {\sqrt {2u}}}
F({5 \over 6}, {1\over 6}, 1;-{1\over u^3}) \nonumber\\
{df_D \over du}&=&{{\sqrt 2}\over 4}u^2 \{ {1\over 6}
F({5 \over 6},{5 \over 6},2;1+u^3)+{5\over 6}
F({11 \over 6},{5 \over 6},2;1+u^3) \}.
\label{sixviii}
\eea
With (\ref{sixv})-(\ref{sixviii}) we can compute the gauge couplings for  
the Higgs variables $a^{(h)}=a$, $a_D^{(h)}=a_D$ and then use the monodromy 
transformations (\ref{xviii}) to obtain the couplings for the variables 
associated to the BPS states becoming massless. We then have:

i) $(0,1)$ monopole:

$$
\left(\begin{array}{c}a^{(m)}_{D} \\
                    a^{(m)} \end{array}\right) =
\left(\begin{array}{cc} 0 &1 \\
                         -1&0\end{array} \right) 
\left(\begin{array}{c}a^{(h)}_D\\
               a^{(h)} \end{array}\right),
$$
$$
\tau_{11}^{(m)}=-{1 \over \tau_{11}^{(h)}}, \,\,\,\,\,\,\,\,\,\
\tau_{01}^{(m)}=-{\tau^{(h)}_{01} \over \tau_{11}^{(h)}},
$$
\be
\tau_{00}^{(m)}=\tau_{00}^{(h)}-{(\tau^{(h)}_{01})^2 \over \tau_{11}^{(h)}}.
\label{sixix}
\ee

ii) $(1,1)$ dyon:

$$
\left(\begin{array}{c}a^{(d)}_{D} \\
                    a^{(d)} \end{array}\right) =
\left(\begin{array}{cc} 0 &-1 \\
                         1&1\end{array} \right) 
\left(\begin{array}{c}a^{(h)}_D\\
               a^{(h)} \end{array}\right),
$$
$$
\tau_{11}^{(d)}=-{1 \over \tau_{11}^{(h)}+1}, \,\,\,\,\,\,\,\,\,\
\tau_{01}^{(d)}={\tau^{(h)}_{01} \over \tau_{11}^{(h)}+1},
$$
\be
\tau_{00}^{(d)}=\tau_{00}^{(h)}-{(\tau^{(h)}_{01})^2 \over \tau_{11}^{(h)}+1}.
\label{sixx}
\ee

iii) $(2,1)$ dyon:

$$
\left(\begin{array}{c}a^{(d_1)}_{D} \\
                    a^{(d_1)} \end{array}\right) =
\left(\begin{array}{cc} 0 &-1 \\
                         1&2\end{array} \right) 
\left(\begin{array}{c}a^{(h)}_D\\
               a^{(h)} \end{array}\right),
$$
$$
\tau_{11}^{(d_1)}=-{1 \over \tau_{11}^{(h)}+2}, \,\,\,\,\,\,\,\,\,\
\tau_{01}^{(d_1)}={\tau^{(h)}_{01} \over \tau_{11}^{(h)}+2},
$$
\be
\tau_{00}^{(d_1)}=\tau_{00}^{(h)}-{(\tau^{(h)}_{01})^2 \over \tau_{11}^{(h)}+2}.
\label{sixxi}
\ee
 
For the other description of the $(2,1)$ dyon we have similar expressions.

Before studying numerically the effective potential, we can extract some 
qualitative information about the vacuum structure and the condensates 
as in sect. 4. As in the $N_f=0$, $SU(2)$ case \cite{soft}, we expect that
 the minima of the effective potential will be located near the singularities 
(the $N=1$ points). The monopole condensate at $u_1$ is given essentially 
by ${\rm Im}\tau^{(m)}_{01}(u_1)$, and we can evaluate 
the $(1,1)$ and $(2,1)$ condensates at the other singularities using the 
${\bf Z}_3$ symmetry in the moduli space, as in (\ref{fourx}). From 
(\ref{sixvii}) we obtain
\be
\tau^{(m)}_{01}(u_1) \sim {du \over da^{(m)}} = 
-2 {\sqrt 2}{\rm e}^{-2 \pi i/3},
\label{sixxii}
\ee
with a non-zero imaginary part, hence we have monopole condensation at 
$u_1={\rm e}^{- \pi i/3}$. The ${\bf Z}_3$ symmetry gives
\be
\tau^{(d)}_{01}(u_2)=-{\rm e}^{ \pi i/3}\tau^{(m)}_{01}(u_1), \,\,\,\,\,\,\
\tau^{(d_1)}_{01}(u_3)=-{\rm e}^{ 2\pi i/3}\tau^{(m)}_{01}(u_1), 
\label{sixxiii}
\ee
and we get a condensate at $u_2={\rm e}^{\pi i/3}$ with the 
same size than the one at $u_1$, 
and no condensate at $u_3=-1$. Therefore we expect 
two minima for the effective 
potential, located near $u_1$ and $u_2$. This is consistent with 
the value of the effective potential at the singularities, 
given by (\ref{fourv}). As $b_{00}\sim -{\rm Re}u$, 
the singularity at $u_3$ 
is energetically less favourable than the singularities at $u_1$, $u_2$, 
which have moreover the same energy. Again we find a precise correlation 
between the size of the condensate and the corresponding value of the 
effective potential.

Now we proceed to the precise numerical analysis 
of the vacuum structure. In the Higgs region the effective 
potential is given by the cosmological constant,
\be
V^{(h)}=-{{\rm det}b^{(h)} \over b^{(h)}_{11}}f_0^2,
\label{higgs}
\ee
which is 
a monodromy invariant. 
\begin{figure}
\centerline{
\hbox{\epsfxsize=6cm\epsfbox{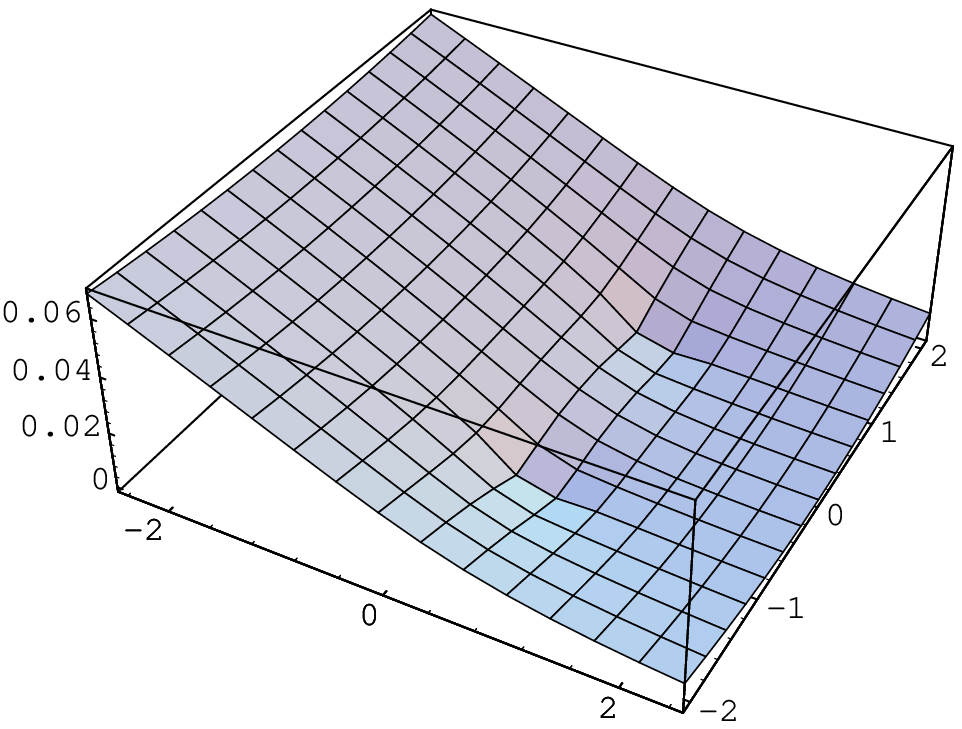}} }
\caption[]{Effective potential $V^{(h)}$, (\ref{higgs}).}
\figlabel\vach
\end{figure}
and we plot it in \fig\vach. Its shape is 
very similar to the $N_f=0$ case studied in \cite{soft}. 
It has two cusp singularities located at $u_1$, $u_2$, where we expect 
an important contribution from the condensates. In fact, one should 
include the monopole and dyon contribution in order to smooth out the 
behavior of the effective potential near these singularities, and 
consider 
\be
V_{\rm eff}=-{{\rm det}b\over b_{11}}f_0^2
-{2 \over b^{(m)}_{11}}\rho^2_{(m)}  
-{2 \over b^{(d)}_{11}}\rho^2_{(d)}.
\label{sixxvii}
\ee
This is 
precisely the expected behaviour from the Wilsonian point of view: 
near the points where extra states become massless one should include 
the relevant degrees of freedom to have a non-singular effective potential. 

Therefore, we must analyze the condensates around the three singularities.
For the $(2,1)$ (or $(-1,1)$) dyon state we find, according to our previous 
estimates in (\ref{sixxiii}), a very tiny condensate, with the same 
pattern obtained for the $(-1,1)$ dyon in the pure Yang-Mills case 
analyzed in \cite{soft}. In particular it does not produce a minimum in the 
effective potential once its contribution is included. 
\begin{figure}
\centerline{
\hbox{\epsfxsize=6cm\epsfbox{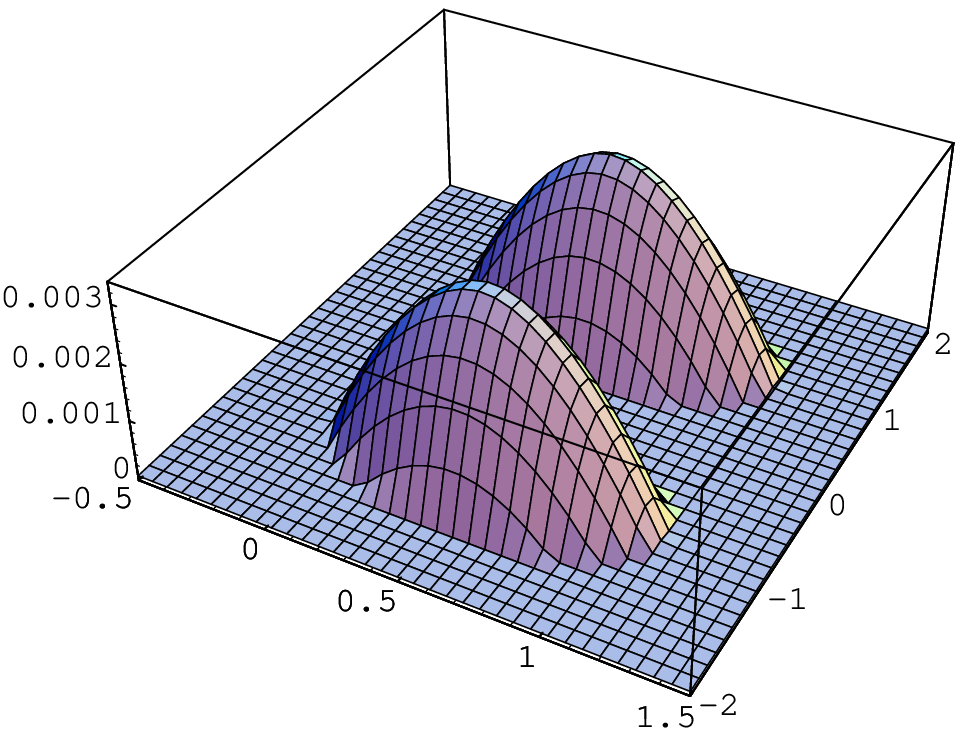}} }
\caption[]{Monopole and dyon VEVs for $f_0=0.1$ on the $u$-plane.}
\figlabel\cond
\end{figure}
For the $(0,1)$ 
monopole and $(1,1)$ dyon associated to the singularities
 at $u_1$, $u_2$, respectively, we find condensates in symmetric regions 
with respect to the ${\rm Re}u$ axis, as one can see in 
\fig\cond. As long as $f_0 <0.25$ the two regions don't 
overlap. Once these contributions are 
included in the effective potential, for this range of $f_0$,
 we get two degenerate minima at
 complex conjugate points $u_0$, $u^{*}_0$. The minimum associated to the 
monopole condensate has ${\rm Im}u_0<0$, and the one associated to the 
dyon condensate has ${\rm Im}u^{*}_0>0$.
\begin{figure}
\centerline{
\hbox{\epsfxsize=6cm\epsfbox{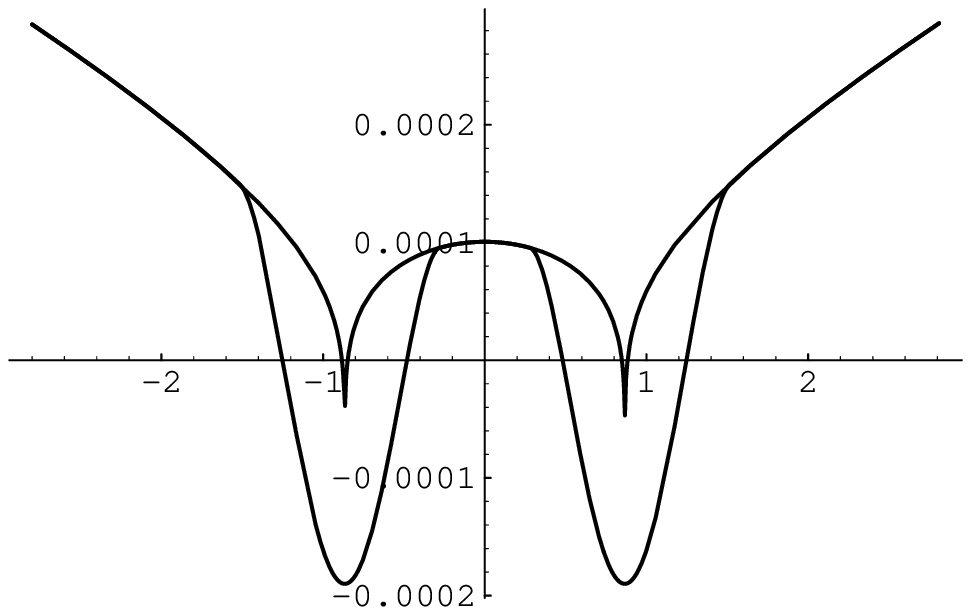}}\qquad
\hbox{\epsfxsize=6cm\epsfbox{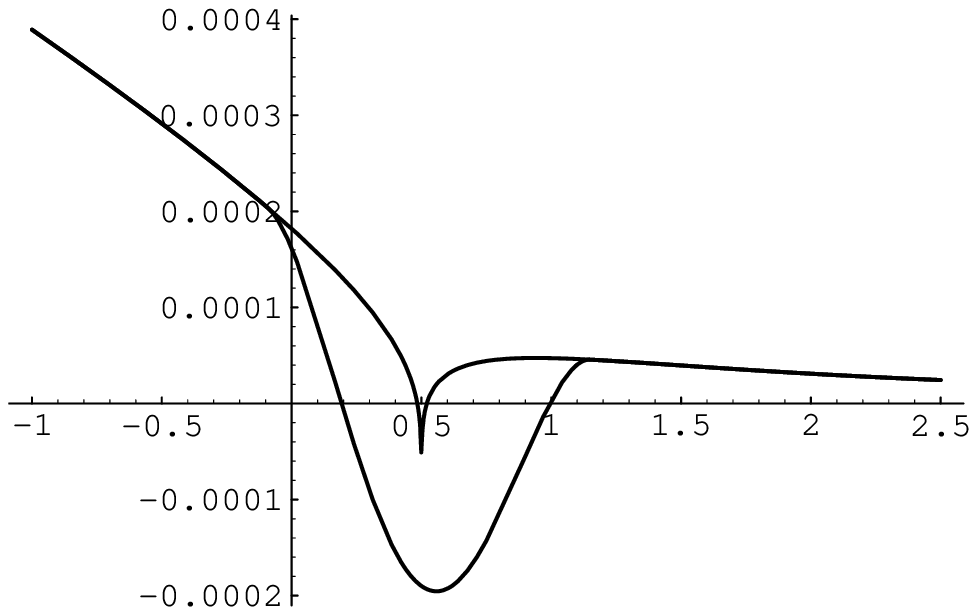}} }
\caption[]{Effective potential (\ref{higgs}) (top),  
and (\ref{sixxvii}) (bottom) for $u=1/2+i y$ (left) and for 
$u=x-i {\sqrt 3}/2$ (right). Both are plotted for
$f_0=0.1$.}
\figlabel\smooth
\end{figure}

Of course, once the contribution 
of the condensates is taken into account, the singularities are smoothed 
out, as one can see in \fig\smooth. The position of the minimum moves 
away from 
the singularities as the supersymmentry breaking parameter $f_0$ is turned on.
 
There are two features which are worth noticing. The symmetry between the 
monopole and dyon condensates, as well as the degeneracy of the potential, 
are due to the following symmetry properties of the monopole and dyon 
variables and couplings:
$$
a^{(m)}(u)={\overline a}^{(d)}(u^*),\,\,\,\,\,\,\,\ 
\tau_{11}^{(m)}(u)=-{\overline \tau}_{11}^{(d)}(u^*),
$$
\be
\tau_{01}^{(m)}(u)={\overline \tau}_{01}^{(d)}(u^*),\,\,\,\,\,\,\,\ 
\tau_{00}^{(m)}(u)=-{\overline \tau}_{00}^{(d)}(u^*).
\label{sixxiv}
\ee
With our choice of the monopole and dyon coordinates, $\epsilon^{(m)}=-1$, 
$\epsilon^{(d)}=1$. From (\ref{sixxiv}), (\ref{fivexv}) and (\ref{fivexvii}) 
one can check that the mass spectrum is the same for the monopole and 
dyon minima. 

The other interesting property of these minima is the following. As it was 
shown in \cite{dyon}, magnetic monopoles labeled by the quantum numbers 
$(n_e, n_m)$ have anomalous electric charge given by 
$q=n_e+(\theta_{\rm eff}/ \pi)n_m$ (in the normalization of 
\cite{swtwo}). The theta parameter we must take into 
account in this case is the low-energy one, 
and as we have been labeling 
the BPS states with quantum numbers referred 
to the Higgs variables $a$, $a_D$, 
it will be given by \cite{swone}
\be
\theta_{\rm eff}= \pi {\rm Re}{da_D \over da}.
\label{sixxv}
\ee
In the $N_f=0$ case analyzed in \cite{soft}, the minimum of the effective 
potential was associated to the $(0,1)$ state and occurred along the real 
$u$-axis with $u>1$. In this region $\theta_{\rm eff}=0$, $q=0$, and we have 
stricitly speaking monopole condensation and confinement of colour electric 
charge, as it happens when the theory is broken down to $N=1$ theory
and the minima are locked at the singularities \cite{konishi}. 
In the softly broken $N_f=1$ theory the 
situation is different: for the minimum associated to the $(0,1)$ BPS 
state one has $\theta_{\rm eff}(u_0)<0$ as soon as the 
supersymmetry breaking parameter is different from zero 
(at the singularity $\theta_{\rm eff}(u_1)=0$), and for the 
minimum located at 
$u_0^*$, associated to the $(1,1)$ BPS state, one has 
$\theta_{\rm eff}(u^*_0)=-1-\theta_{\rm eff}(u_0)$ (which in fact 
is a consequence of the symmetry (\ref{sixxiv})).  
 In this way the monopole 
and dyon 
have anomalous electric charges given by
\be
q^{(m)}={\theta_{\rm eff}(u_0)\over \pi}, \,\,\,\,\,\,\ 
q^{(d)}=-{\theta_{\rm eff}(u_0)\over \pi}.
\label{sixxvi}
\ee
As the supersymmetry breaking parameter is turned on, the minimum 
moves in such a way that $|\theta_{\rm eff}(u_0)|$ increases starting 
from zero : the 
condensing states have greater electric charges with opposite sign. 
Hence we have dyon condensation properly speaking in both minima. This 
must produce some screening of the electric sources and correspondingly 
a smaller string tension, although on general grounds we still expect 
confinement of colour electric charge. 

As $f_0 \sim 0.25$, the regions where monopole and dyon condensation occur 
begin to overlap on the real $u$-axis. This kind of behaviour for the 
softly broken $N=2$ theories was already noticed in \cite{soft}. At this 
point it is clear that in the overlapping region there are non-mutually 
local degrees of freedom which must be taken into account simultaneously,
 and as it is 
well known it is difficult to find an effective lagrangian description 
of this situation. Nevertheless it is interesting to notice that we are 
studying the vacuum structure with the effective potential (\ref{sixxvii}),
 and in fact 
we still have a smooth description of the low-energy physics when we 
add the cosmological constant and the condensates contribution, even in the 
overlapping region. 
\begin{figure}
\centerline{
\hbox{\epsfxsize=6cm\epsfbox{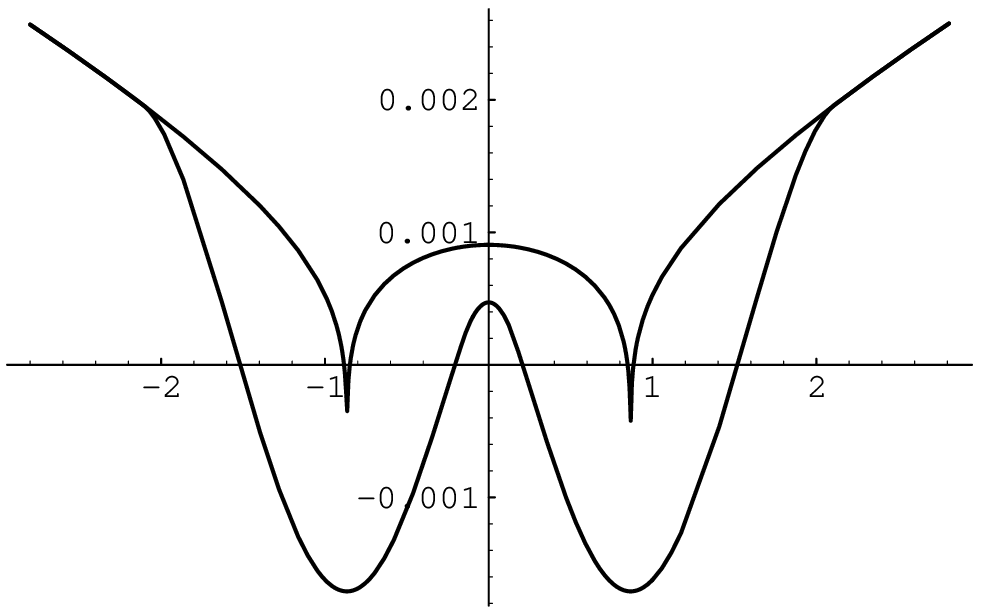}} }
\caption[]{Effective potential (\ref{sixxvii}) for $f_0=0.3$ 
along $u=1/2 + i y$.}
\figlabel\potnl
\end{figure}
In \fig\potnl\ $V_{\rm eff}$ is plotted along the ${\rm Im}u$ 
direction, for ${\rm Re}u=1/2$, and we can see that 
the overlapping of monopole and dyon condensates lowers 
the energy on the real $u$-axis. A reliable description in terms of 
(\ref{sixxvii}) is possible because 
of the monodromy invariance of the cosmological 
constant, which takes into account the gauge-field degrees of freedom 
independently of the description we choose. In the contributions coming from 
the BPS states we are not considering the monopole and dyon variables as 
independent variables, as the only independent parameter in
 (\ref{sixxvii}) is the gauge-invariant order parameter $u$. 
Hence we can try to extrapolate our analysis to a wider range of the 
supersymmetry breaking parameter using the effective potential 
description. This could give a hint of the dynamics 
induced by the interaction of non-mutually local objects in field theory. 
The situation we find here is similar to the one considered by Argyres 
and Douglas in \cite{ds}, concerning certain points in the 
moduli space of the 
$N=2$, $SU(3)$ Yang-Mills theory where non-local BPS states become 
simultaneously massless. They also showed that, even if it is not possible 
to write a well-defined lagrangian describing these objects, one can make 
sense of other quantities such as the $\beta$-function. We think that 
(\ref{sixxvii}) should be considered on the same footing.     
    
However, the monopole and dyon condensates appearing in (\ref{sixxvii}) are 
defined in terms of functions which have branch cuts connecting the origin 
of the moduli space with the other singularities. In other words, the terms 
associated to the BPS states in (\ref{sixxvii}) are not monodromy invariant. 
The monopole (dyon) condensate attains these branch cuts precisely when 
the overlapping occurs, and one could think that this invalidates the 
effective potential description as soon as $f_0 \sim 0.25$. But actually one 
can solve this problem performing an analytic continuation through the cuts. 
This is closely related to the fact that inside the curve of marginal 
stability the 
BPS states are described by different quantum numbers, depending 
on the region under consideration \cite{bf}. Finally, 
this also indicates what is 
the breakdown of our approximation: as soon as the monopole (dyon) 
condensate attains the other singularities, the analytic continuation cannot 
be done in a consistent way. Moreover, the condensates have a non-smooth  
behaviour at the singularities associated to the other, non-mutually local 
BPS states. This is an indication that we are not taking into account the 
relevant degrees of freedom in the description provided by (\ref{sixxvii}). 
This breakdown occurs for $f_0 \sim 0.8$, therefore we can study the 
dynamics induced by the supersymmetry breaking parameter before it 
reaches this critical value. 

For $0.25 \le f_0 \le 0.6$ the vacuum structure remains qualitatively the same, 
with the only difference that the minima of the effective potential move 
away from the singularities and correspondingly $|\theta_{\rm eff}(u_0)|$ 
increases its value. For $f_0 \sim 0.6$, a new minimum develops on the real 
$u$-axis, due to the overlapping of the monopole and dyon VEVs. If we increae 
$f_0$ we find a first order phase transition for 
$f_0 \sim 0.68$: the new minimum becomes the absolute minimum 
and therefore the true vacuum of the 
theory, and the vacuum degeneracy is lifted. 
\begin{figure}
\centerline{
\hbox{\epsfxsize=6cm\epsfbox{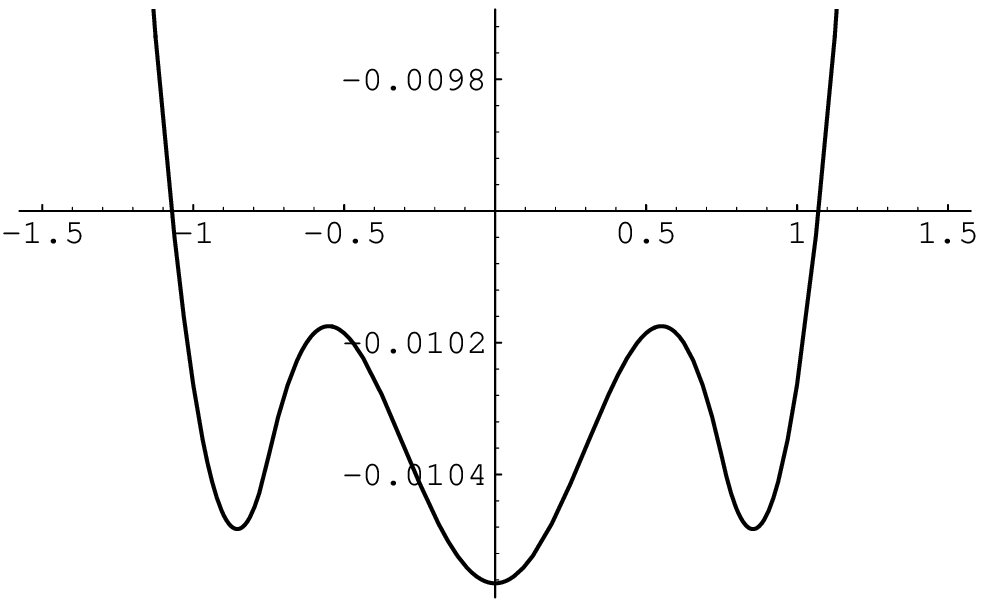}}\qquad
\hbox{\epsfxsize=6cm\epsfbox{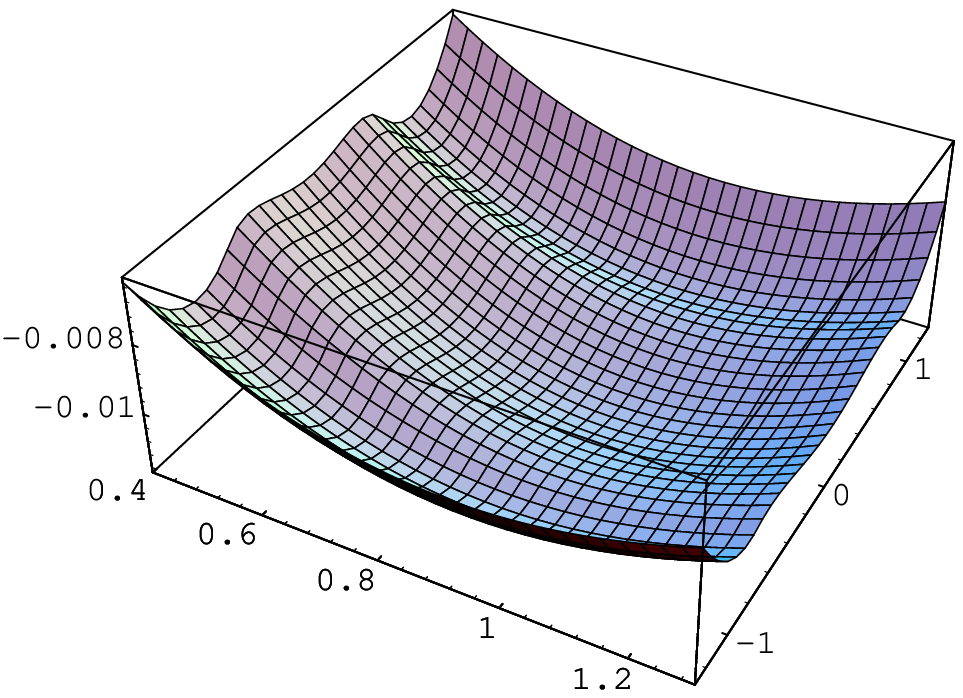}} }
\caption[]{Effective potential (\ref{sixxvii}) 
for $u=0.95+i y$, $f_0=0.68$ (left), and for 
$f_0=0.7$ on the $u$-plane (right).}
\figlabel\newmin
\end{figure}
In \fig\newmin\ $V_{\rm eff}$ is plotted for $f_0=0.68$ along the 
${\rm Im}u$ direction with ${\rm Re}u=0.95$, and for $f_0=0.7$ on the 
$u$-plane. The possibility of this kind 
of behaviour due to the presence in the effective potential of non-mutually 
local states was pointed out in \cite{soft}. This minimum persists until 
the region where the monopole (dyon) develops 
a VEV attains the other singularities. At this point our description  
breaks down and we can not longer trust the effective potential. 

What can be the interpretation of this phase transition? A possible 
explanation may be the fact that the BPS states which are condensing have an 
anomalous electric charge due to the effective theta parameter. One 
should expect that the additional electric charge makes the condensation 
less and less favourable energetically, as it happens in many models. 
The dynamics looks very much like the phase transition in the 
theta angle leading to an oblique confinement phase. In these cases the 
anomalous electric charge makes more favourable enegetically the condensation 
of a bound state of dyons with opposite electric charges \cite{ob}. In our 
model, the theta angle transition is induced by the supersymmetry breaking 
parameter $f_0$, and the new minimum seems to correspond to a simultaneous 
condensation of dyons which also have opposite electric charges: the 
theta parameter along the real $u$-axis is $\theta_{\rm eff}=\pi/2$, and 
the monopole and dyon have anomalous electric charge $q^{(m)}=-q^{(d)}=-1/2$.  

It is clear that a thorough understanding of this kind of vacua, 
with non-mutually local states, is still lacking. But we think that the 
analysis presented here can give some hints about the possibility of 
these new phases and the rich dynamics associated to them. Perhaps the 
transition to this new vacuum can be understood in more traditional terms 
as the condenstion of a bound state with zero electric charge, 
but the ubiquity of these 
phenomena in supersymmetric theories \cite{ad,apsw} raises the 
possibility of new phases of gauge theories that may be relevant to 
the description of the QCD vacuum.

\newpage

{\large\bf Acknowledgements}

We would like to thank Mike Douglas for a useful discussion, 
and the Departamento de F\'\i sica Te\'orica at Universidad 
Aut\'onoma de Madrid, where part of this work was done, for its hospitality. 
M.M. would also like to thank the Theory 
Division at CERN
for its hospitality. 
The work of M.M.~is supported in part by DGICYT under grant
PB93-0344 and by
CICYT under grant AEN94-0928.

\end{document}